\documentclass[10pt,aps,prl,twocolumn,superscriptaddress,preprintnumbers,floatfix,notitlepage]{revtex4-1}
\usepackage{hyperref}
\hypersetup{
    colorlinks=true,
    linkcolor=blue,
    filecolor=magenta,      
    urlcolor=blue,
}
\usepackage{amsmath,amsfonts,amssymb}
\usepackage{fancyhdr}
\usepackage{graphicx}
\usepackage{xspace}
\usepackage{rotating}
\usepackage[normalem]{ulem}
\usepackage{braket}
\usepackage{verbatim}
\usepackage{xcolor}
\usepackage{hyperref}
\usepackage[utf8]{inputenc}
\usepackage{slashed}

\def\lsim{\mathrel{\raise.3ex\hbox{$<$\kern-.75em\lower1ex\hbox{$\sim$}}}}
\def\gsim{\mathrel{\raise.3ex\hbox{$>$\kern-.75em\lower1ex\hbox{$\sim$}}}}

\newcommand{\be}{\begin{equation}}
\newcommand{\ee}{\end{equation}}
\newcommand{\bea}{\begin{equation}\begin{aligned}}
\newcommand{\eea}{\end{aligned}\end{equation}}

\begin{document}


\title{Dilatonic Couplings and the Relic Abundance of Ultralight Dark Matter}

\author{Ahmad Alachkar}
\affiliation{Theoretical Particle Physics and Cosmology, King's College London, Strand, London, WC2R 2LS, United Kingdom}


\author{Malcolm Fairbairn}
\affiliation{Theoretical Particle Physics and Cosmology, King's College London, Strand, London, WC2R 2LS, United Kingdom}

\author{David J. E.~Marsh}
\affiliation{Theoretical Particle Physics and Cosmology, King's College London, Strand, London, WC2R 2LS, United Kingdom}

\begin{abstract}
Models of scalar field dark matter where the scalar is a dilaton have a special behaviour, since non-trivial couplings, $d$, to matter result in a contribution to the potential for the field which is proportional to the trace of the stress-energy tensor.  We look in more detail at the dilaton mass, $m_\phi$, and initial conditions required to yield the correct relic abundance for couplings that are not already excluded by terrestrial experiments. In minimal models with only couplings accessible to terrestrial searches, we find that dilaton dark matter with $m_{\phi} \gtrsim 10^{-10}$  eV requires couplings suppressed compared to constraints from equivalence principle (EP) tests and fifth force searches in order to not produce too much dark matter, improving on the strongest current experimental constraints by up to $\sim {\cal O}(10)$, with consequences for the proposed mechanical resonator dilaton DM searches.  In non-minimal or universally coupled models, the unconstrained couplings of the dilaton to e.g. the top quark can strongly influence the relic abundance at all masses. In particular, this implies that atom interferometry searches at masses $m_\phi\approx 10^{-19}\text{ eV}$ are unable to constrain the early Universe behaviour or UV physics of the dilaton. We also find that dilatonic couplings allow for compatibility of $m_\phi \gtrsim 10^{-7}\text{ eV}$ with an observably large tensor-to-scalar ratio in the cosmic microwave background, which is not possible for a decoupled scalar of the same mass. 
\\
~~\\

\end{abstract}

\maketitle



Einstein's theory of general relativity is extremely successful \cite{will2014confrontation,reynaud2008tests,GRtest1,GRtest2}, as is the Standard Model (SM) of particle physics \cite{odom2006new,aad2012observation}.  The combination of the two theories leads to the prediction of a hot big bang with primordial nucleosynthesis \cite{Steigman:2007xt} and Cosmic Microwave Background (CMB) radiation \cite{spergel2003first,aghanim2020planck,akrami2020planck} that are observed to have occurred in our Universe.  Despite these huge successes, multiple independent observations have also indicated the presence of some so far undetected dark matter (DM) \cite{Marsh:2024ury} which  is not part of the Standard Model of particle physics.  Furthermore, the ultraviolet (UV) completion of both gravity and quantum field theory is still a subject of much debate. One leading candidate is string/M-theory. Since string theory contains no dimensionful constants, and requires dimensional reduction to describe our 3+1 dimensional world, the low energy effective field theory (EFT) generically contains a plethora of scalar (and pseudoscalar) degrees of freedom. If these degrees of freedom are light, then they provide candidates for DM, and possible IR modifications of gravity. The possible existence of such particles is the subject of intense experimental study at present~\cite{2168507,arvanitaki2015searching,Adams:2022pbo}. 

General Relativity is obtained by varying the Einstein Hilbert action plus a matter action with canonical kinetic terms. This is referred to as the case of minimal coupling between matter and gravity.  Theories of gravity with non-minimal couplings between matter and curvature, and non-minimal matter kinetic terms, have been explored extensively over the past decades as possible explanations of, or being related to, dark matter \cite{arvanitaki2015searching,banerjee2023phenomenology,brzeminski2021time}, dark energy \cite{brax2010dilaton,khoury2004chameleon} and the UV completion of QFT and gravity \cite{narain1989new,taylor1988dilaton}.

One example of such investigations is the idea of dilatonic dark matter (DDM), where the non-minimal coupling between gravity and the SM is given by a dynamical scalar field, the oscillation of which can provide stress energy compatible with the equation of state of DM. The DDM relic abundance can be provided by a process similar to vacuum realignment of axion DM~\cite{preskill1983cosmology,abbott1983cosmological,dine1983not}, which is specified by the potential of the dilaton, and the initial conditions of the field. (See \cite{hubisz2024note} for a critique of the generation of cosmologically relevant dilatonic dark matter via such mechanisms.)

As with any successful and phenomenologically interesting theory of DM, a connection between the relic abundance and the couplings constrained in the laboratory is desirable if we are to use observations to learn about the underlying fundamental theory of DM, and its connection to early Universe and UV physics. In the following we will show that, on the one hand at very low dilaton masses $m_\phi\approx 10^{-19}\text{ eV}$ there are significant unknowns associated with making such a connection which prevent one from linking direct detection to early Universe physics and the initial conditions of the dilaton in the UV. On the other hand, however, at larger dilaton masses $m_\phi \approx 10^{-7}\text{ eV}$, the unique physics of the dilaton can allow compatibility between DDM and certain models of inflation, which is not possible for generic scalar DM of the same mass.

The low energy EFT of DDM can be modelled by a generalised scalar-tensor theory defined by the following action 
\begin{equation}
\begin{aligned}
    S(g_{\mu\nu},\varphi,\Psi_i) &= \frac{1}{2\kappa^{2}}\int d^4x \sqrt{-g}  \Big[ \varphi R -\frac{\omega}{\varphi} (\nabla \varphi)^{2}  
    \\ &-U(\varphi)  \Big]  + S_{m}(g_{\mu\nu}, \varphi, \Psi_{i}),  
    \label{eq: dilaton action Jordan Frame}
\end{aligned}    
\end{equation}
where $R$ is the Ricci scalar constructed from the metric $g_{\mu\nu}$ with determinant $g$, $\omega=-1$ is a dimensionless coupling constant called the dilatonic Brans-Dicke parameter, $\kappa \equiv \sqrt{8\pi G}$ with $G$ the bare gravitational constant (i.e. the gravitational constant in the absence of the scalar interaction), $U(\varphi)$ is the scalar field potential defined in this conformal frame, referred to hereinafter as the matter frame, $S_m$ is the matter action and $\Psi_{i}$ denotes the matter fields in the SM.  
An equivalent formulation of the theory (modulo boundary terms) is given in the Einstein frame (here and throughout this work, Einstein frame quantities are labelled with a tilde), where the kinetic term for the metric is of the Einstein-Hilbert form, via a Weyl rescaling $ g_{\mu\nu} \longrightarrow \tilde{g}_{\mu\nu}= \Omega^{2}(x){g}_{\mu\nu}$, with $\Omega^2 \equiv F(\varphi) = \varphi \equiv \kappa \phi$, and a redefinition $\tilde{\phi} = \sqrt{\frac{|2\omega+3|}{2\kappa^{2} }} \ln\left(\frac{\phi}{\phi_0}\right)$ such that the field is canonically normalised. $\phi_0$ is chosen to be  $\kappa^{-1}$ so that the scalar field gives the correct value of $G$ today. 

The effective low-energy, matter-gravity-dilaton action in the Einstein frame is then 
\begin{equation}
\begin{aligned}
    S(\tilde{g}_{\mu\nu},\tilde{\phi},\Psi_i)&= \int d^4x \sqrt{-\tilde{g}} \Big[ \frac{1}{2\kappa^{2}} \tilde{R} -\frac{1}{2} \tilde{g}^{\mu\nu}\nabla_{\mu}\tilde{\phi}\nabla_{\nu}\tilde{\phi} \\
    &-V(\tilde{\phi}) \Big]  + S_{m}(\Omega^{-2}\tilde{g}_{\mu\nu},\tilde{\phi}, \Psi_{i}),  
    \label{eq: dilaton action Einstein Frame}
\end{aligned}    
\end{equation}
where $V(\tilde{\phi})= U(\varphi)/(2\kappa^{2}  \varphi^{2})$.  
\begin{figure*}
    \centering

    \includegraphics[width=0.8\textwidth]{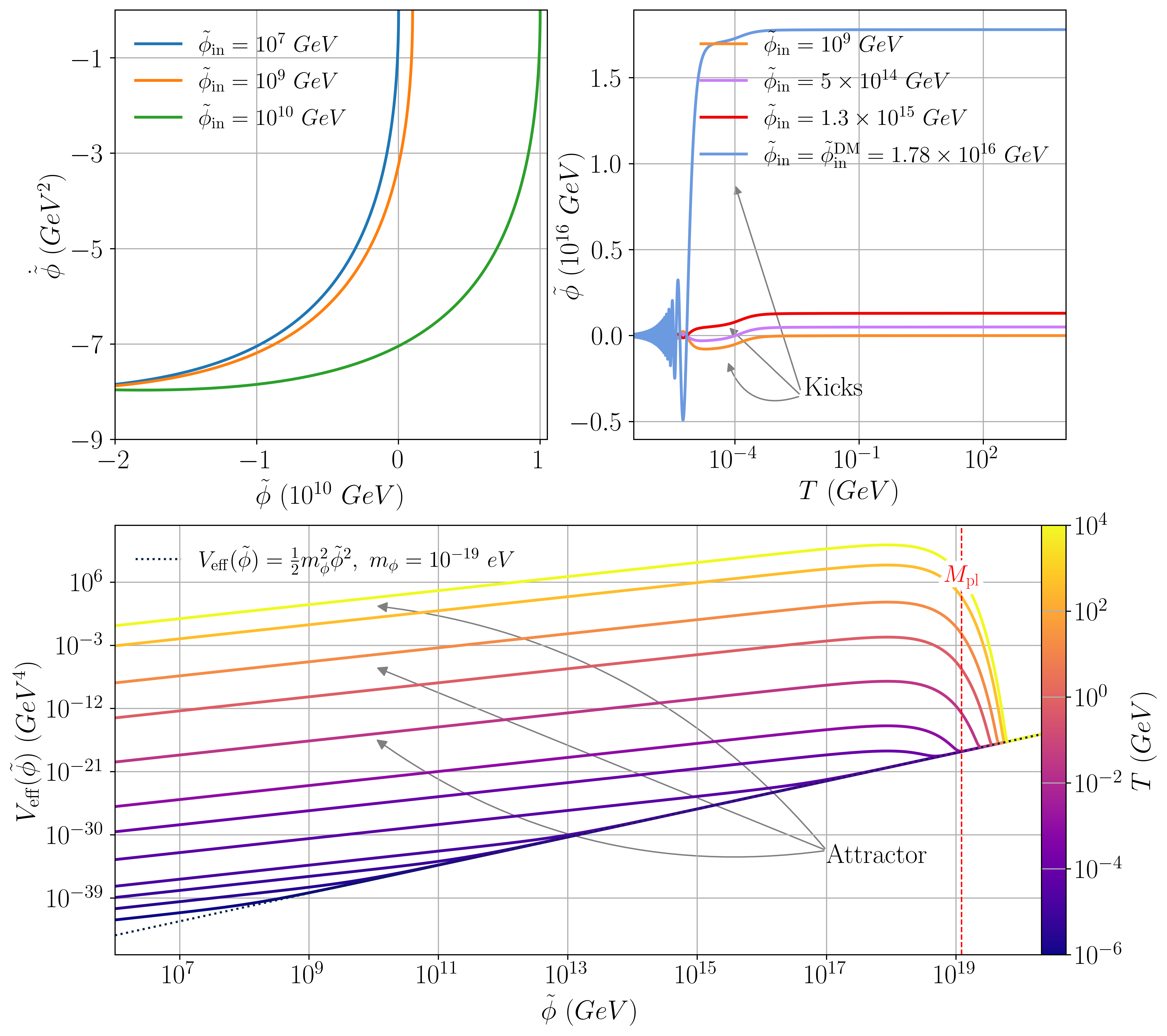}
    \caption{\it \textbf{Bottom}: the effective potential as a function of expectation value for the field for DDM with mass $m_{\phi}=10^{-19}$ eV, where the couplings sensitive to current EP and fifth force constraints are set at their maximum permissible values, for different values of temperature.  The dotted line is the purely harmonic potential. \textbf{Top left}: the phase portrait for different initial field values that give rise to an attractor solution, corresponding to the linear regime in the effective potential. \textbf{Top right}: The cosmological evolution of the dilaton, for the same model parameters, as function of temperature (right), for a range of initial field values (see Fig. \ref{fig:1e-19_UV_sensitivity}), showing the kicks due to the interaction with SM species. $\tilde{\phi}^{\mathrm{DM}}_\mathrm{in}$ is the initial field value that gives the measured DM density today. 
    }
    \label{fig:Effective Potential}
\end{figure*}

\emph{Dilaton evolution and relic abundance:} We solve the dilaton evolution in the Einstein frame effective potential using the Friedmann-Robertson-Walker (FRW) background during the radiation dominated era, with scale factor $\tilde{a}$ and Hubble parameter $\tilde{H}=\dot{\tilde{a}}/\tilde{a}$. When the dilaton mass is smaller than the Hubble scale during inflation, this leads to a uniform initial field value, $\tilde{\phi}_{\mathrm{in}}$, and $d\tilde{\phi}/d\tilde{t}|_{\tilde{t}=\tilde{t}_\mathrm{in}}\approx 0$. The effective potential and evolution of the dilaton is illustrated in Fig.~\ref{fig:Effective Potential}.

The dynamics of the dilaton are governed by the Klein-Gordon equation, $\tilde{\Box}\tilde{\phi}= -\partial V/ \partial \tilde{\phi} - \beta(\tilde{\phi})  \tilde{T}^{\mu}_{~ \mu}  -\tilde{\sigma}$, 
where we  define the Einstein frame scalar charge density of the sources $\tilde{\sigma}$ by the variation with respect to $\tilde{\phi}$ as $\tilde{\sigma} \equiv \frac{1}{\sqrt{-\tilde{g}}}\frac{\delta S_m }{\delta \tilde{\phi}}$~\cite{hees2018violation,campbell2021searching}, and $\tilde{\Box} \equiv \tilde{g}^{\mu\nu} \tilde{\nabla}_{\mu}\tilde{\nabla}_{\nu}$ is the d’Alembert operator for the covariant derivative $\tilde{\nabla}_{\mu}$, with Levi-Civita connection associated with the rescaled metric. The parameter $\beta(\tilde{\phi}) \equiv \frac{1}{2} d\ln F(\phi)/d \tilde{\phi}= \sqrt{\kappa^2/(2(2\omega+3))}$ is a universal coupling that characterizes the departure of this generalised scalar-tensor theory from GR, in the absence of $\tilde{\sigma}$. 
The Einstein frame stress-energy tensor is given by $\tilde{T}_{\mu\nu} \equiv  (-2/\sqrt{-\tilde{g}}) \delta S_{m}/\delta \tilde{g}^{\mu\nu}$, which sources the Friedmann equation:
\begin{equation}\label{eq: Friedmann equation}
3\tilde{H}^{2}= \kappa^{2}\left[\tilde{\rho} + \frac{1}{2}\Big(\frac{d\tilde{\phi}}{d\tilde{t}}\Big)^{2} +V(\tilde{\phi}) \right],
\end{equation}
and the associated (redundant) acceleration equation for $d\tilde{H}/d\tilde{t}$.
We drop the dilaton contribution to the right hand side when computing the relic abundance (DM should be sub-dominant during the radiation era), but note that the dilaton affects $\tilde{\rho}$ implicitly via the matter frame temperature $T$.

In flat FRW the dilaton field equation is:
\begin{equation}\label{eq: KG eq final}
\frac{d^{2}\tilde{\phi}}{d \tilde{t}^{2}}  
 +3\tilde{H}\frac{d \tilde{\phi}}{d \tilde{t}} + \frac{d V}{d \tilde{\phi}} =\beta(\tilde{\phi}) \sum_i \Big(-\tilde{\rho}_{i}+3\tilde{p}_{i} \Big) -\tilde{\sigma},
\end{equation}
where the index $i$ spans the various Standard Model components of the Universe (hadrons, radiation, etc.) with corresponding densities $\rho_{i}$ and pressures $p_{i}$. The parameter $\tilde{\sigma}$ is given by
\begin{equation}
\begin{split}
\tilde{\sigma} &= 
\left(\kappa^2 \phi_{0}^{2} \exp\left(2\sqrt{\frac{2\kappa^{2}}{2\omega+3}}\tilde{\phi}\right)\right)^{-1} \Bigg[\kappa d_{m_e} (\rho_{e}-3p_{e}) \\ 
&+ \kappa d_{m_u} (\rho_{u}-3p_{u}) 
+ \kappa d_{m_d} (\rho_{d}-3p_{d}) + \kappa d_e\frac{ \pi^2}{3}\frac{\alpha}{4\pi} T^4  \Bigg].
\end{split}
\end{equation}
We include the contribution to the induced dilaton mass from thermal field theory~\cite{bouley2023constraints}, where the leading diagram appears at two loops. 

We consider a simple harmonic potential to stabilise the dilaton given by $V(\tilde{\phi})=m_{\phi}^2\tilde{\phi}^2/2$, and 
determine the dynamics by numerically solving the coupled system of equations comprising Eq. (\ref{eq: Friedmann equation}) and (\ref{eq: KG eq final}).  This system of equations is closed when we specify how the matter-frame temperature $T$ depends on $\tilde{a}$ and $\tilde{\phi}$, following the prescription of \cite{erickcek2014chameleons}. 
We evaluate the number of degrees of freedom, $g_{\ast S}(T)$ for the SM particle
spectrum, obtained from \cite{saikawa2018primordial}, treating the QCD phase transition as instantaneous at a temperature of $T_{\mathrm{QCD}} = 150$ MeV.

At early times deep in the radiation era, $\tilde{H}^{2}\gg m^{2}_{\mathrm{eff}}$, where $m_\text{eff}$ is the effective dilaton mass given by the total potential, damping any initial velocity possessed by the field. We compute the evolution of $\tilde{\phi}$ for a range of masses $m_{\phi}$, initial conditions $ \tilde{\phi}_\mathrm{in}$ and coupling strengths $d_i$ and determine the relic abundance using a WKB approximation to redshift to the present once oscillations are within the harmonic regime. The evolution of the dilaton can generally be split into three regimes depending on which of the following three terms dominates Eq.~(\ref{eq: KG eq final}): the Hubble friction term $\tilde{H}$, the bare mass $m_{\phi}$ or the induced mass due to $\beta(\tilde{\phi})  \tilde{T}^{\mu}_{~ \mu}  +\tilde{\sigma}$.

The effective potential, $V_{\mathrm{eff}}$, governing the dynamics, plotted in Fig \ref{fig:Effective Potential} for $m_{\phi}= 10^{-19}$ eV with SM couplings set to existing EP and fifth force constraints (see below for definition), consists of the zero temperature quadratic term (bare mass) as well as a temperature-dependent linear term due to the trace of stress-energy tensor (induced potential), exponentially suppressed by $2\sqrt{\frac{2\kappa^{2}}{2\omega+3}}\tilde{\phi}$ due to the conformal mapping of physical quantities in the matter frame. 
The field value of $\tilde{\phi}^{0}_{\mathrm{in}}= 1/(2\sqrt{\frac{2\kappa^{2}}{2\omega+3}}\tilde{\phi}) \sim 9\times 10^{17}$ GeV separates two regimes for fixed $T \gtrsim 5 \times 10^{4}$ eV. At $\tilde{\phi}_{\mathrm{in}}>\tilde{\phi}^{0}_{\mathrm{in}}$, the field rolls down $V_{\mathrm{eff}}$ towards larger values initially before the bare potential dominates, where $\Delta \tilde{\phi} <0$, and $\tilde{\phi}$ henceforth exhibits the standard behaviour of a damped (by $\tilde{H}$) harmonic oscillator. For $\tilde{\phi}_{\mathrm{in}} \leq \tilde{\phi}^{0}_{\mathrm{in}}$, we have $\Delta \tilde{\phi} <0$  throughout the evolution, along the constant-slope induced potential $\propto \ln F(\phi(\tilde{\phi}))$. Now fixing $\tilde{\phi}_{\mathrm{in}}$, for $T < 5 \times 10^{4}$ eV , the field excursion is always negative, towards $\tilde{\phi}_{\mathrm{min}}$ which coincides with the bare potential minimum at low enough $T$ (late times). 

For a radiation fluid, the trace of the stress-
energy tensor is zero during most of the early 
Universe. However, as the radiation bath cools, 
whenever $T \sim m_i$ for a SM species with rest 
mass $m_i$ there is a non-zero contribution for 
about one e-fold of expansion as the particle $i$ 
becomes non-relativistic and its pressure decreases 
at a faster rate than its energy density. This 
generates a `kick' in the forcing term imparting a 
significant velocity to the dilaton- an effect first 
studied in \cite{damour1993tensor,damour1994string}, 
in the context of scalar-tensor models. For 
$\tilde{\phi}_{\mathrm{in}} \leq 
\tilde{\phi}^{0}_{\mathrm{in}}$ these kicks, along 
with the linear part of the potential, give rise to 
an attractor solution. This is demonstrated in the 
phase portrait shown in top left 
Fig.~\ref{fig:Effective Potential}. The kicks are 
visible in the field evolution shown in the top right of Fig.~\ref{fig:Effective Potential}. \\

\emph{Minimal Fifth-Force Consistent Couplings.} We 
have seen above how the couplings between the 
dilaton and SM fields modify the effective potential 
of the dilaton, leading to attractor behaviour in 
the evolution of the field. Direct detection of DDM 
relies on these same couplings, and so it is natural 
to ask to what level the dilaton evolution is 
affected by observably large values for the 
couplings. We consider the following couplings to 
the light fields of the SM (at leading order), which 
are those probed, e.g., by atom interferometers: 
\clearpage
\begin{equation} \label{eq: interacting lagrangian}
\begin{aligned}
\mathcal{L}&=\mathcal{L}_{\mathrm{SM} }+\mathcal{L}_{\mathrm{int} \phi}\\
&=\mathcal{L}_{\mathrm{SM} }+\kappa \tilde{\phi} \Big[+\frac{d_e}{4 e^2} F_{\mu \nu} F^{\mu \nu}-\frac{d_g \beta_3}{2 g_3} F_{\mu \nu}^A F^{A \mu \nu} \\
&-\sum_{i=e, u, d}\left(d_{m_i}+\gamma_{m_i} d_g\right) m_i \bar{\psi}_i \psi_i \Big],
\end{aligned}
\end{equation}
introducing linear interaction terms to the effective
Lagrangian describing the physics of ground state nuclei, $\mathcal{L}_{\mathrm{SM} }$. $F_{\mu\nu}$ is the electromagnetic Faraday tensor, $F_{\mu\nu}^{A}$ is the gluon field strength tensor, $\beta_3$ is the $\beta$-function for the
running of $g_3$, $m_i$ is the mass of the fermions, $\gamma_{m_i}$ is the anomalous dimension giving the energy running of the
masses of the quarks, and $\psi_{i}$ denotes the Weyl fermion spinors. A $\tilde{\phi}$-dependent coupling to the
kinetic term of the fermion, $f(\tilde{\phi}) \bar{\psi_i} i \slashed{D} \psi_i$, is omitted as it can be absorbed in a
suitable $\tilde{\phi}$–dependent rescaling of $\psi_i$. 

At low energies of  $\lesssim 1$ GeV, one has integrated out the effect of weak interactions, and the heavy quarks $c$, $b$ and $t$ (see \cite{damour2010equivalence} for a discussion on the
issue of $\tilde{\phi}$ sensitivity of effects linked to the
strange quark). 
This leaves us with the electron $e$,
the $u$ quark, and the $d$ quark, with interactions mediated
by the electromagnetic ($A_\mu$) and gluonic ($A^{A}_\mu$) fields. 

Up to linear order, each of the five terms in $\mathcal{L}_{\mathrm{SM}}$ can couple to $\kappa \tilde{\phi}$ via a dimensionless coefficient, $d_i$. 
Our parameterisation of the couplings $d_i$ is relative to gravity~\cite{damour2010equivalence}, so $d_{i}=1 $ would be the universal couplings of a scalar graviton. This normalisation implies that the physical meaning of the five dilaton-coupling coefficients ${d_i}$ is to introduce a $\tilde{\phi}$-dependence (and hence temporal variation) in the
parameters entering the low-energy physics of the form $X(\tilde{\phi}) =\left(1+d_i \kappa \tilde{\phi} \right) X$, where $X \in (\Lambda_3, \alpha, m_{j})$, corresponding to $i \in (g, e, j)$ in that order, for $j= e, u, d$, $\alpha$ denotes the fine structure constant and $\Lambda_3$ is the QCD mass scale.

\begin{figure}
    \centering
    \includegraphics[width=0.48\textwidth]{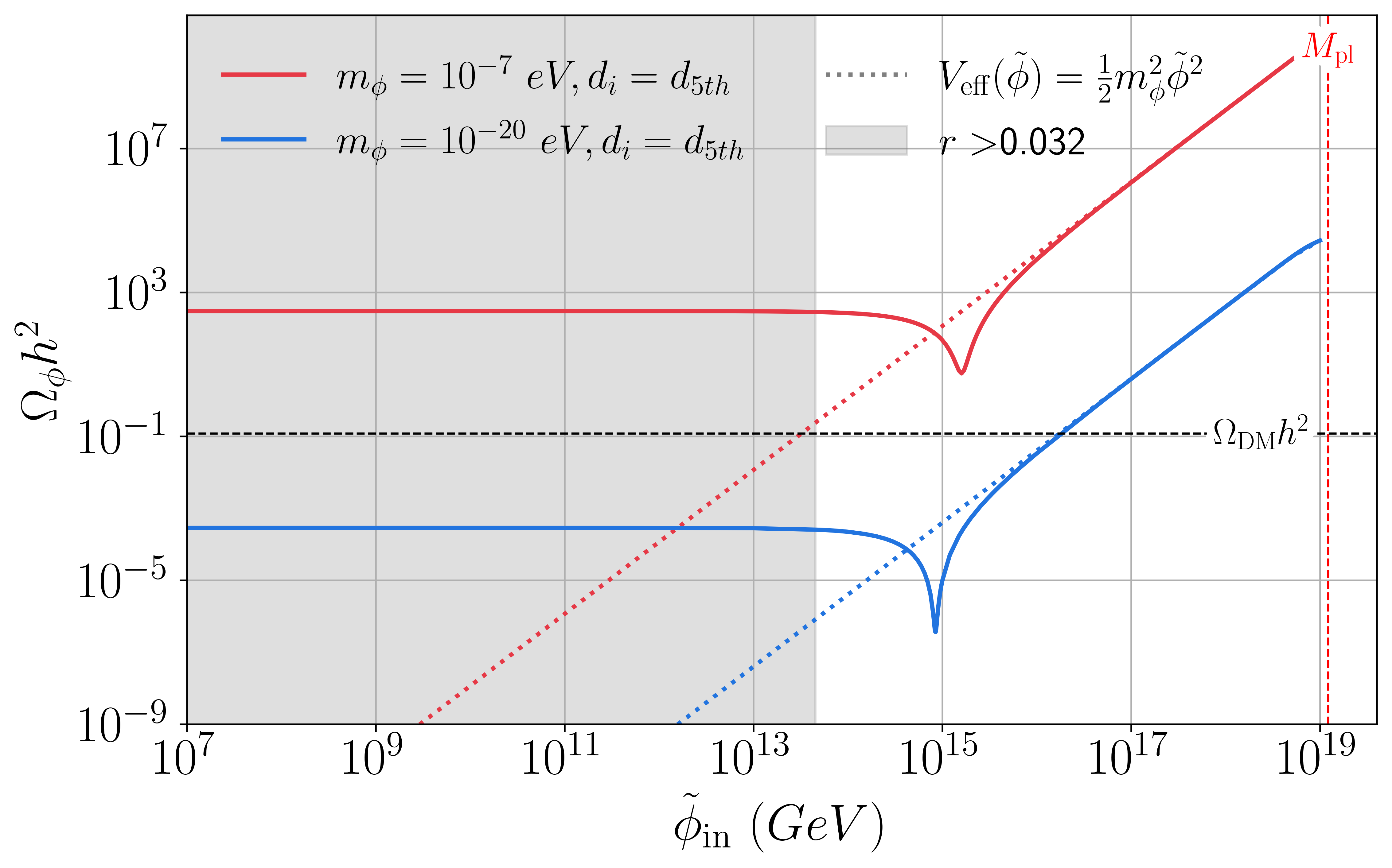}
    \caption{\it Relic abundance vs initial value of scalar for masses at the two opposite ends of the range investigated. Solid lines show the case of the dilaton with minimal EP and fifth-force consistent couplings, while dotted lines show the case of a purely harmonic potential. The grey region would be excluded for hypothetical detection of tensor-to-scalar ratio at the current upper limit $r=0.032$.}
    \label{fig:1e-7_minimal_Hinf}
\end{figure}
\begin{figure}
    \centering
    \includegraphics[width=0.48\textwidth]{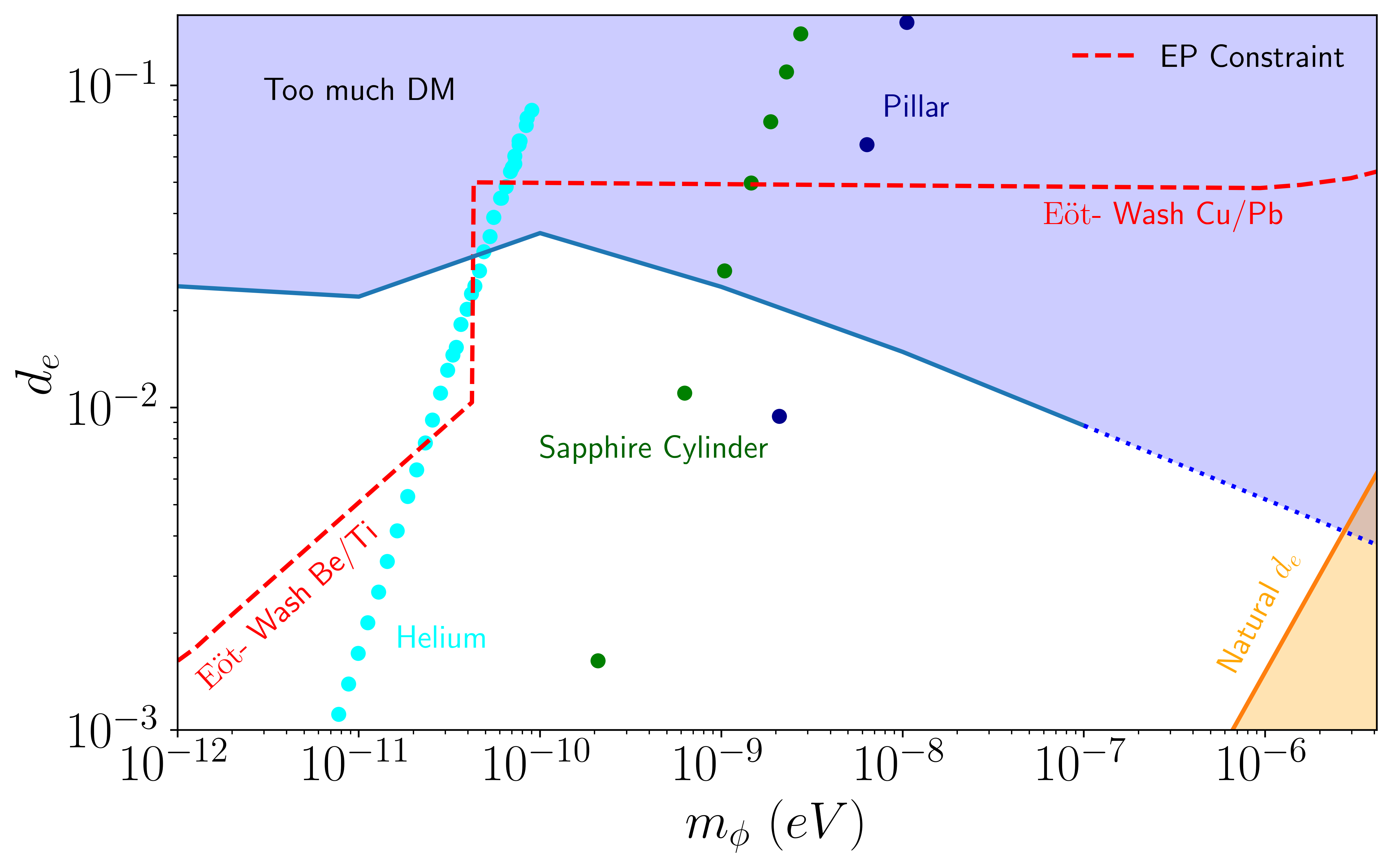}
    \caption{\it Maximum coupling strength $d_{e}$ corresponding to the correct DM relic abundance vs. dilaton mass $m_{\phi}$, with the other three couplings of the minimal model set at the maximum permissible values $d_{5th}$, given the existing EP and fifth force constraints. The dotted line is an extrapolation to higher masses, which are not accessible by our simulations. Overlaid are current experimental constraints on $d_e$ given by the E$\ddot{o}$t-Wash equivalence principle test. The solid circles are projections for the minimum detectable coupling for three compact mechanical resonators proposed as candidate detectors for scalar dark matter \cite{manley2020searching}: a superfluid helium bar resonator (light blue) \cite{de2017ultra}, a cylindrical HEM\textsuperscript{\textregistered} sapphire test mass \cite{rowan2000investigation}  (green) and a sapphire micropillar (dark blue) resonator \cite{Neuhaus}. The orange region is natural for coupling to the fine structure constant with a 10 TeV cutoff \cite{manley2020searching,arvanitaki2016sound}. }
    \label{fig:d_e_constraints}
\end{figure}

For a minimal scenario where only the above couplings are present, not just in the IR but at all scales, we fix the couplings to have the largest values allowed by EP and fifth force constraints~\cite{wagner2012torsion,schlamminger2008test,smith1999short,berge2018microscope,banerjee2023phenomenology,bouley2023constraints}, and switch off couplings to all other SM species. We then calculate the dilaton relic abundance as a function of initial field value $\tilde{\phi}_{\mathrm{in}}$, for a range of masses $m_{\phi} \in \left[10^{-22},  10^{-7}\right]$ eV, which we illustrate with two examples in Fig.~\ref{fig:1e-7_minimal_Hinf}.

We find that lower masses (in the range $10^{-22}-10^{-12}$ eV) give solutions for the relic abundance for $\tilde{\phi}_{\mathrm{in}}$ that are very close to the case of a quadratic potential (zero dilatonic couplings), whereas higher masses give solutions that deviate strongly from this. This is because for the relatively low values of $\tilde{\phi}_\mathrm{in}$ required at high mass, the relic density  stays constant and the system's behaviour is relatively invariant over a range of initial conditions due to the scalar field attractor, leading to the relic abundance fixed only by $m_\phi$ (holding couplings fixed). At the large initial field values required at low dilaton mass, the induced potential is sub-dominant to the bare potential and $\Omega_{\phi} h^{2} \propto \tilde{\phi}_{\mathrm{in}}^{2}$, giving solutions that align with that of a purely quadratic potential. Thus the early Universe behaviour of minimal DDM at low mass is insensitive to the exact values of the couplings in the minimal model, once they are below EP and fifth force constraints. 

On the other hand, DDM at larger masses $m_{\phi} \gtrsim 10^{-10}\text{ eV}$ requires couplings suppressed with respect to their EP and fifth force values in order to give a consistent relic abundance, due to small field values and the dominance of the attractor. This is illustrated in more detail in Fig.~\ref{fig:d_e_constraints}, where we compute the maximum allowed value of $d_e$ (the coupling to the fine structure constant, which has the dominant effect on the cosmological evolution) required in order to not produce too much DM. At the largest mass we have considered, $m_{\phi}=10^{-7}\text{ eV}$, $d_e$ should be suppressed by a factor of 10 compared to the EP constraints. Extrapolating our numerical results to higher dilaton masses $m_{\phi} \gtrsim 10^{-6}\text{ eV}$ the constraint from DM overproduction reaches the `natural' values of couplings expected with 10 TeV UV cut-off~\cite{manley2020searching,arvanitaki2016sound}, which is a regime defined by the criterion that loop corrections to the dilaton mass are less than the physical mass $m_{\phi}$~\cite{dimopoulos1996macroscopic}.

\begin{figure}
    \centering
    \includegraphics[width=0.48\textwidth]{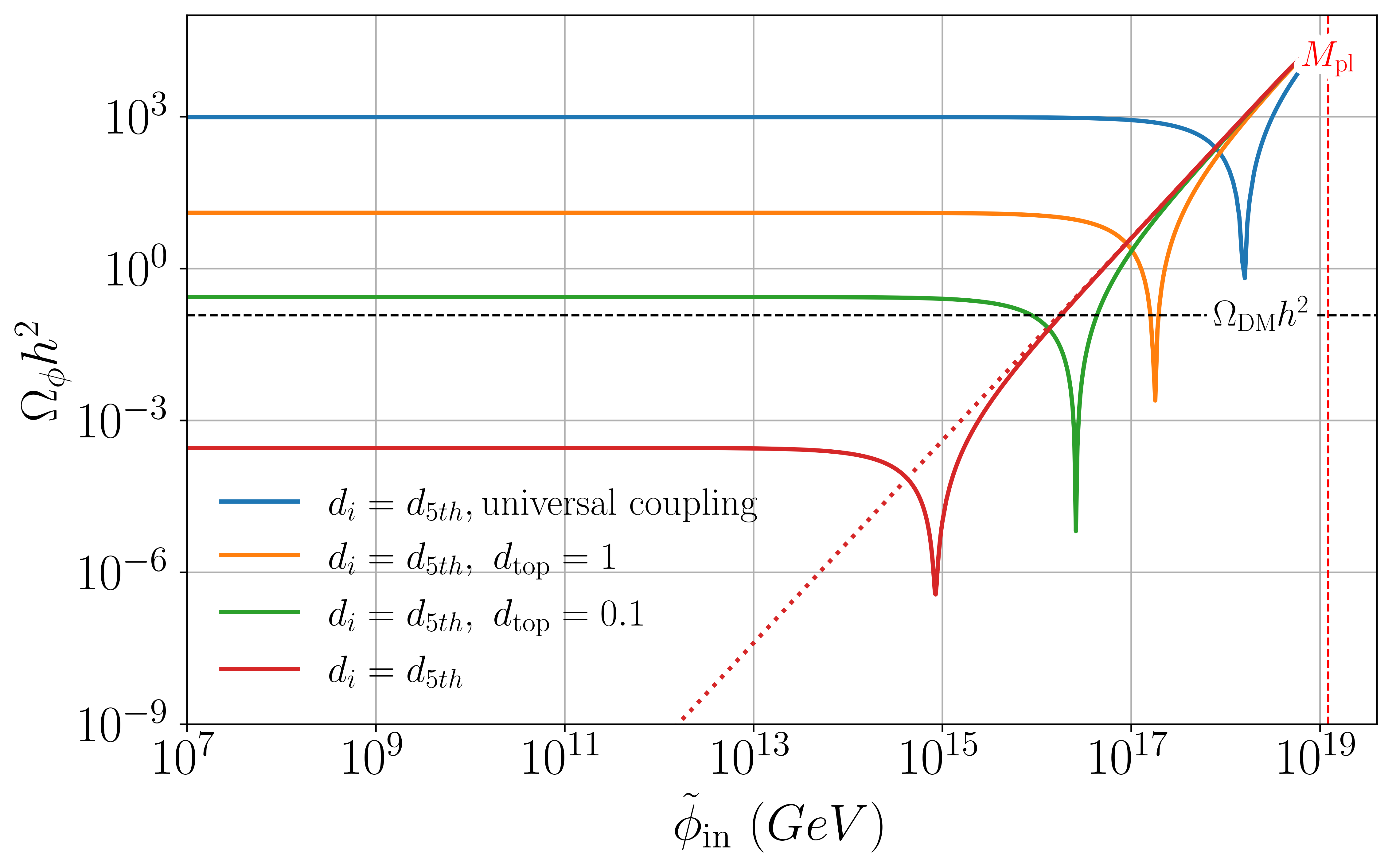}
    \caption{\it Four different scenarios all with DDM with mass $m_\phi=10^{-19}$ eV where all the couplings sensitive to current EP and fifth force constraints are set at their maximum permissible values, denoted by $d_{5th}$. The dotted line again denotes the vanilla model with completely minimal coupling to gravity (no dilatonic couplings to SM), corresponding to a purely harmonic potential. Any slight variation in the unconstrained coupling to top quarks or indeed a universal coupling of unity to all SM particles not probed by EP or fifth force experiments greatly change the value of $\tilde{\phi}_{\mathrm{in}}$ to obtain good relic abundance ($i=u,q,e,\gamma$).}
    \label{fig:1e-19_UV_sensitivity}
\end{figure}

\emph{The Case of Unconstrained Couplings:} 
We now consider the effect of the coupling of the dilaton to other SM species that are not constrained experimentally, for example to the top quark. 
This is shown for $m_{\phi}= 10^{-19}$ eV in Fig.~\ref{fig:1e-19_UV_sensitivity}, where in addition to the minimal model couplings at EP and fifth force constraints level, we add a coupling $d_{\mathrm{top}}$ at 0.1 and 1, or a universal coupling to all SM species. 
If such an unconstrained $d$ is switched on, we find 
different solutions for the relic abundance 
$\Omega_{\phi}h^{2}$ as a function of the initial 
conditions $\tilde{\phi}_\mathrm{in}$ and mass 
$m_{\phi}$. We thus conclude that, without a strong 
reason to exclude dilaton couplings to the heavy SM 
fields, it is generally impossible to make 
inferences about the early Universe behaviour and UV 
physics of DDM based on results informed only by 
existing experiments. 

\emph{Relation to models of inflation:} Given the precise measurement of the scalar amplitude of the CMB~\cite{aghanim2020planck}, the value of the tensor-to-scalar ratio, $r$, determines the Hubble parameter during inflation, $H_{\mathrm{\ast}}$, as $H_{\ast}= 8 \times 10^{13} \sqrt{r/0.1}$ GeV.
The existing upper limit on $r$ is $r<0.032$ at $95 \%$ confidence \cite{tristram2022improved,ade2021improved,collaboration2020planck}.

As a spectator field, the expectation value of the DDM will be excited during inflation, carrying out a random walk with rms amplitude $\Theta\simeq\sqrt{N} H_{\ast}/(2 \pi)$ \cite{enqvist2012spectator}, where $N$ is the number of efolds, which we fix to 60. A higher energy scale during inflation will result in a greater expectation value for $\tilde{\phi}_{\mathrm{in}}$, ruling out scenarios where relic abundance requires lower $\tilde{\phi}_{\mathrm{in}}$. The grey region on Fig.~\ref{fig:1e-7_minimal_Hinf} represents the region of parameter space which would be ruled out if $r=0.032$, i.e. measured near the current upper limit. We observe that scalar DM with no dilatonic couplings (dashed line) and $m_\phi>10^{-7}\text{ eV}$ would produce too much relic abundance and thus cannot compose the DM in this case (note that a vanilla scalar field dark matter model with completely minimal coupling to gravity would be also in tension with a value of $r=0.032$, c.f. Ref.~\cite{Marsh:2014qoa}). The introduction of dilatonic couplings can tune the location and size of the minimum in the relic abundance at larger $\tilde{\phi}_{\mathrm{in}}$, potentially leading to a scenario where scalar DM with $m_\phi>10^{-7} \text{ eV}$ is consistent with high scale inflation. 

\emph{Conclusion:} Our study of the DDM relic abundance has two main implications for DM direct detection at relatively high and relatively low dilaton mass. Firstly, at low masses, probed for example by AION~\cite{badurina2020aion}, the existence of unconstrained couplings to unstable higher mass particles which do not contribute to EP limits prevents a completely final conclusion being drawn on the order of magnitude of $\tilde{\phi}_{\mathrm{in}}$ based on a detection of DDM with fixed mass. This means, in contrast to the case of the axion for example, it is slightly more complicated to connect DDM direct detection to initial conditions in the Early Universe. At high masses, the DDM relic abundance is affected by the dilatonic couplings and the energy scale of inflation. Simultaneous measurement of scalar DM and a large tensor-to-scalar ratio in the CMB, corresponding to high scale inflation, is inconsistent if the scalar DM has no interactions beyond $m_\phi^2\phi^2$, but can be made consistent by introducing dilatonic couplings. In the happy event of this dual detection, it is possible to make inferences about the presently unconstrained dilatonic couplings necessary for consistency. Lastly, we showed how the coupling to the fine structrue constant can be constrained by the requirement that DM is not overabundant, improving on existing limits by up to an order of magnitude, probing natural values of the coupling, and with consequences for mechanical resonator DDM searches~\cite{manley2020searching}.

\section*{Acknowledgements}
  
The work of AA was supported by a UK STFC studentship,
that of MF was supported in part by STFC Grants ST/P000258/1 and ST/T000759/1.
The work of DJEM is supported by STFC grant ST/T004037/1.
We are grateful to Edmund Copeland for discussions and correspondence, and to Angelo Maggi for collaboration at an early stage of this work.


\bibliographystyle{unsrt}
\bibliography{prl}

\begin{thebibliography}{10}

\bibitem{will2014confrontation}
Clifford~M Will.
\newblock The confrontation between general relativity and experiment.
\newblock {\em Living reviews in relativity}, 17(1):1--117, 2014.

\bibitem{reynaud2008tests}
Serge Reynaud and Marc-Thierry Jaekel.
\newblock Tests of general relativity in the solar system.
\newblock {\em arXiv preprint arXiv:0801.3407}, 2008.

\bibitem{GRtest1}
Stephan Schlamminger, K-Y Choi, Todd~A Wagner, Jens~H Gundlach, and Eric~G Adelberger.
\newblock Test of the equivalence principle using a rotating torsion balance.
\newblock {\em Physical Review Letters}, 100(4):041101, 2008.

\bibitem{GRtest2}
Robert~FC Vessot, Martin~W Levine, Edward~M Mattison, EL~Blomberg, TE~Hoffman, GU~Nystrom, BF~Farrel, Rudolph Decher, Peter~B Eby, CR~Baugher, et~al.
\newblock Test of relativistic gravitation with a space-borne hydrogen maser.
\newblock {\em Physical Review Letters}, 45(26):2081, 1980.

\bibitem{odom2006new}
Brian Odom, David Hanneke, Brian D’Urso, and Gerald Gabrielse.
\newblock New measurement of the electron magnetic moment using a one-electron quantum cyclotron.
\newblock {\em Physical review letters}, 97(3):030801, 2006.

\bibitem{aad2012observation}
Georges Aad, Tatevik Abajyan, Brad Abbott, Jalal Abdallah, S~Abdel Khalek, Ahmed~Ali Abdelalim, R~Aben, B~Abi, M~Abolins, OS~AbouZeid, et~al.
\newblock Observation of a new particle in the search for the standard model higgs boson with the atlas detector at the lhc.
\newblock {\em Physics Letters B}, 716(1):1--29, 2012.

\bibitem{Steigman:2007xt}
Gary Steigman.
\newblock {Primordial Nucleosynthesis in the Precision Cosmology Era}.
\newblock {\em Ann. Rev. Nucl. Part. Sci.}, 57:463--491, 2007.

\bibitem{spergel2003first}
David~N Spergel, Licia Verde, Hiranya~V Peiris, Eiichiro Komatsu, MR~Nolta, Charles~L Bennett, Mark Halpern, Gary Hinshaw, Norman Jarosik, Alan Kogut, et~al.
\newblock First-year wilkinson microwave anisotropy probe (wmap)* observations: determination of cosmological parameters.
\newblock {\em The Astrophysical Journal Supplement Series}, 148(1):175, 2003.

\bibitem{aghanim2020planck}
Nabila Aghanim, Yashar Akrami, Mark Ashdown, J~Aumont, C~Baccigalupi, M~Ballardini, AJ~Banday, RB~Barreiro, N~Bartolo, S~Basak, et~al.
\newblock Planck 2018 results-vi. cosmological parameters.
\newblock {\em Astronomy \& Astrophysics}, 641:A6, 2020.

\bibitem{akrami2020planck}
Yashar Akrami, Kristian~Joten Andersen, Mark Ashdown, Carlo Baccigalupi, M~Ballardini, Anthony~J Banday, RB~Barreiro, Nicola Bartolo, S~Basak, K~Benabed, et~al.
\newblock Planck intermediate results-lvii. joint planck lfi and hfi data processing.
\newblock {\em Astronomy \& Astrophysics}, 643:A42, 2020.

\bibitem{Marsh:2024ury}
David J.~E. Marsh, David Ellis, and Viraf~M. Mehta.
\newblock {\em {Dark Matter: Evidence, Theory, and Constraints}}.
\newblock Princeton University Press, 9 2024.

\bibitem{2168507}
Derek F.~Jackson Kimball and Karl van Bibber, editors.
\newblock {\em {The Search for Ultralight Bosonic Dark Matter}}.

\bibitem{arvanitaki2015searching}
Asimina Arvanitaki, Junwu Huang, and Ken Van~Tilburg.
\newblock Searching for dilaton dark matter with atomic clocks.
\newblock {\em Physical Review D}, 91(1):015015, 2015.

\bibitem{Adams:2022pbo}
C.~B. Adams et~al.
\newblock {Axion Dark Matter}.
\newblock In {\em {Snowmass 2021}}, 3 2022.

\bibitem{banerjee2023phenomenology}
Abhishek Banerjee, Gilad Perez, Marianna Safronova, Inbar Savoray, and Aviv Shalit.
\newblock The phenomenology of quadratically coupled ultra light dark matter.
\newblock {\em Journal of High Energy Physics}, 2023(10):1--62, 2023.

\bibitem{brzeminski2021time}
Dawid Brzeminski, Zackaria Chacko, Abhish Dev, and Anson Hook.
\newblock Time-varying fine structure constant from naturally ultralight dark matter.
\newblock {\em Physical Review D}, 104(7):075019, 2021.

\bibitem{brax2010dilaton}
Philippe Brax, Carsten van~de Bruck, Anne-Christine Davis, and Douglas Shaw.
\newblock Dilaton and modified gravity.
\newblock {\em Physical Review D}, 82(6):063519, 2010.

\bibitem{khoury2004chameleon}
Justin Khoury and Amanda Weltman.
\newblock Chameleon fields: Awaiting surprises for tests of gravity in space.
\newblock {\em Physical review letters}, 93(17):171104, 2004.

\bibitem{narain1989new}
Kumar~S Narain.
\newblock New heterotic string theories in uncompactified dimensions< 10.
\newblock In {\em Current Physics--Sources and Comments}, volume~4, pages 246--251. Elsevier, 1989.

\bibitem{taylor1988dilaton}
TR~Taylor and Gabriele Veneziano.
\newblock Dilaton couplings at large distances.
\newblock {\em Physics Letters B}, 213(4):450--454, 1988.

\bibitem{preskill1983cosmology}
John Preskill, Mark~B Wise, and Frank Wilczek.
\newblock Cosmology of the invisible axion.
\newblock {\em Physics Letters B}, 120(1-3):127--132, 1983.

\bibitem{abbott1983cosmological}
Laurence~F Abbott and P~Sikivie.
\newblock A cosmological bound on the invisible axion.
\newblock {\em Physics Letters B}, 120(1-3):133--136, 1983.

\bibitem{dine1983not}
Michael Dine and Willy Fischler.
\newblock The not-so-harmless axion.
\newblock {\em Physics Letters B}, 120(1-3):137--141, 1983.

\bibitem{hubisz2024note}
Jay Hubisz, Shaked Ironi, Gilad Perez, and Rogerio Rosenfeld.
\newblock A note on the quality of dilatonic ultralight dark matter.
\newblock {\em arXiv preprint arXiv:2401.08737}, 2024.

\bibitem{hees2018violation}
Aur{\'e}lien Hees, Olivier Minazzoli, Etienne Savalle, Yevgeny~V Stadnik, and Peter Wolf.
\newblock Violation of the equivalence principle from light scalar dark matter.
\newblock {\em Physical Review D}, 98(6):064051, 2018.

\bibitem{campbell2021searching}
William~M Campbell, Ben~T McAllister, Maxim Goryachev, Eugene~N Ivanov, and Michael~E Tobar.
\newblock Searching for scalar dark matter via coupling to fundamental constants with photonic, atomic, and mechanical oscillators.
\newblock {\em Physical Review Letters}, 126(7):071301, 2021.

\bibitem{bouley2023constraints}
Thomas Bouley, Philip S{\o}rensen, and Tien-Tien Yu.
\newblock Constraints on ultralight scalar dark matter with quadratic couplings.
\newblock {\em Journal of High Energy Physics}, 2023(3):1--25, 2023.

\bibitem{erickcek2014chameleons}
Adrienne~L Erickcek, Neil Barnaby, Clare Burrage, and Zhiqi Huang.
\newblock Chameleons in the early universe: kicks, rebounds, and particle production.
\newblock {\em Physical Review D}, 89(8):084074, 2014.

\bibitem{saikawa2018primordial}
Ken'ichi Saikawa and Satoshi Shirai.
\newblock Primordial gravitational waves, precisely: The role of thermodynamics in the standard model.
\newblock {\em Journal of Cosmology and Astroparticle Physics}, 2018(05):035, 2018.

\bibitem{damour1993tensor}
Thibault Damour and Kenneth Nordtvedt.
\newblock Tensor-scalar cosmological models and their relaxation toward general relativity.
\newblock {\em Physical Review D}, 48(8):3436, 1993.

\bibitem{damour1994string}
Thibault Damour and Alexander~M Polyakov.
\newblock The string dilation and a least coupling principle.
\newblock {\em Nuclear Physics B}, 423(2-3):532--558, 1994.

\bibitem{damour2010equivalence}
Thibault Damour and John~F Donoghue.
\newblock Equivalence principle violations and couplings of a light dilaton.
\newblock {\em Physical Review D}, 82(8):084033, 2010.

\bibitem{manley2020searching}
Jack Manley, Dalziel~J Wilson, Russell Stump, Daniel Grin, and Swati Singh.
\newblock Searching for scalar dark matter with compact mechanical resonators.
\newblock {\em Physical review letters}, 124(15):151301, 2020.

\bibitem{de2017ultra}
LA~De~Lorenzo and KC~Schwab.
\newblock Ultra-high \uppercase{Q} acoustic resonance in superfluid$^{4}$ \uppercase{H}e.
\newblock {\em Journal of Low Temperature Physics}, 186:233--240, 2017.

\bibitem{rowan2000investigation}
S~Rowan, G~Cagnoli, P~Sneddon, J~Hough, R~Route, EK~Gustafson, MM~Fejer, and V~Mitrofanov.
\newblock Investigation of mechanical loss factors of some candidate materials for the test masses of gravitational wave detectors.
\newblock {\em Physics Letters A}, 265(1-2):5--11, 2000.

\bibitem{Neuhaus}
L.~Neuhaus.
\newblock Ph.\uppercase{D}. thesis.
\newblock {\em Université Pierre et Marie CurieParis VI}, 2016.

\bibitem{arvanitaki2016sound}
Asimina Arvanitaki, Savas Dimopoulos, and Ken Van~Tilburg.
\newblock Sound of dark matter: searching for light scalars with resonant-mass detectors.
\newblock {\em Physical review letters}, 116(3):031102, 2016.

\bibitem{wagner2012torsion}
Todd~A Wagner, S~Schlamminger, JH~Gundlach, and Eric~G Adelberger.
\newblock Torsion-balance tests of the weak equivalence principle.
\newblock {\em Classical and Quantum Gravity}, 29(18):184002, 2012.

\bibitem{schlamminger2008test}
Stephan Schlamminger, K-Y Choi, Todd~A Wagner, Jens~H Gundlach, and Eric~G Adelberger.
\newblock Test of the equivalence principle using a rotating torsion balance.
\newblock {\em Physical Review Letters}, 100(4):041101, 2008.

\bibitem{smith1999short}
GL~Smith, CD~Hoyle, JH~Gundlach, EG~Adelberger, Blayne~R Heckel, and HE~Swanson.
\newblock Short-range tests of the equivalence principle.
\newblock {\em Physical Review D}, 61(2):022001, 1999.

\bibitem{berge2018microscope}
Joel Berg{\'e}, Philippe Brax, Gilles M{\'e}tris, Martin Pernot-Borr{\`a}s, Pierre Touboul, and Jean-Philippe Uzan.
\newblock Microscope mission: first constraints on the violation of the weak equivalence principle by a light scalar dilaton.
\newblock {\em Physical review letters}, 120(14):141101, 2018.

\bibitem{dimopoulos1996macroscopic}
S~Dimopoulos and Gian~Francesco Giudice.
\newblock Macroscopic forces from supersymmetry.
\newblock {\em Physics Letters B}, 379(1-4):105--114, 1996.

\bibitem{tristram2022improved}
Matthieu Tristram, Anthony~J Banday, Krzysztof~M G{\'o}rski, Reijo Keskitalo, CR~Lawrence, Kristian~Joten Andersen, R~Bel{\'e}n Barreiro, J~Borrill, LPL Colombo, HK~Eriksen, et~al.
\newblock Improved limits on the tensor-to-scalar ratio using bicep and p l a n c k data.
\newblock {\em Physical Review D}, 105(8):083524, 2022.

\bibitem{ade2021improved}
Peter~AR Ade, Z~Ahmed, M~Amiri, D~Barkats, R~Basu Thakur, CA~Bischoff, D~Beck, JJ~Bock, H~Boenish, E~Bullock, et~al.
\newblock Improved constraints on primordial gravitational waves using planck, wmap, and bicep/keck observations through the 2018 observing season.
\newblock {\em Physical review letters}, 127(15):151301, 2021.

\bibitem{collaboration2020planck}
Planck Collaboration et~al.
\newblock Planck intermediate results-lvii. joint planck lfi and hfi data processing.
\newblock {\em A\&A}, 643:A42, 2020.

\bibitem{enqvist2012spectator}
Kari Enqvist, Rose~N Lerner, Olli Taanila, and Anders Tranberg.
\newblock Spectator field dynamics in de sitter and curvaton initial conditions.
\newblock {\em Journal of Cosmology and Astroparticle Physics}, 2012(10):052, 2012.

\bibitem{Marsh:2014qoa}
David J.~E. Marsh, Daniel Grin, Renee Hlozek, and Pedro~G. Ferreira.
\newblock {Tensor Interpretation of BICEP2 Results Severely Constrains Axion Dark Matter}.
\newblock {\em Phys. Rev. Lett.}, 113(1):011801, 2014.

\bibitem{badurina2020aion}
L~Badurina, E~Bentine, Diego Blas, K~Bongs, D~Bortoletto, T~Bowcock, K~Bridges, W~Bowden, O~Buchmueller, C~Burrage, et~al.
\newblock Aion: an atom interferometer observatory and network.
\newblock {\em Journal of Cosmology and Astroparticle Physics}, 2020(05):011, 2020.

\end{thebibliography}

\end{document}